\documentclass[letterpaper,twoside]{jpconf}

\usepackage{fancyhdr}
\usepackage{iopams}
\usepackage{graphicx}

\def\Ghat{\hat G}
\def\Rhat{\hat R}
\def\Vhat{\hat V}
\def\Psihat{\hat \Psi}
\def\epshat{\hat \epsilon}
\def\tauhat{\hat \tau}

\def\mo{\mathaccent"17 m}

\def\pdot{\dot p}
\def\qdot{\dot q}
\def\taudot{\dot \tau}
\def\zetadot{\dot \zeta}
\def\pbdotdot{{\mathbf {\,\ddot {\! \mbox{\it p}}}}}

\def\dl{{[\hskip -1.5pt [}}         
\def\dr{{]\hskip -1.5pt ]}}         

\def\degree{\mathaccent"17 {}}

\begin{document}

\title{Quantum effects from a purely geometrical relativity theory\footnote{
Presented at the VI Mexican School on Gravitation and Mathematical Physics
``Approaches to Quantum Gravity'', Playa del Carmen, Quintana Roo, Mexico,
November 21--27, 2005.}}

\author{Homer G Ellis}

\address{Department of Mathematics, University of Colorado at Boulder,
  395 UCB, Boulder, Colorado 80309, USA}

\ead{Homer.Ellis@Colorado.EDU}

\begin{abstract}
A purely geometrical relativity theory results from a construction that
produces from three-dimensional space a happy unification of Kaluza's
five-dimensional theory and Weyl's conformal theory.  The theory can provide
geometrical explanations for the following observed phenomena, among others:
(a) lifetimes of elementary particles of lengths inversely proportional to
their rest masses; (b)~the equality of charge magnitude among all charged
particles interacting at an event; (c)~the propensity of electrons in atoms to
be seen in discretely spaced orbits; and (d)~`quantum jumps' between those
orbits.  This suggests the possibility that the theory can provide a
deterministic underpinning of quantum mechanics like that provided to
thermodynamics by the molecular theory of gases.
\end{abstract}

This presentation is intended to show that some of the phenomena thought to be
explainable only through the process of `quantizing' a classical relativistic
theory can be explained, qualitatively and to some degree quantitatively, by a
purely geometrical relativity theory based on and derived solely from the
geometry of three-dimensional space.\footnote{An early, imperfect description
of the theory can be found in \cite{1}, later, detailed descriptions in
\cite{2}.} A simple geometric construction applied iteratively generates new
dimensions beyond the basic three of space.  The first application produces
space-time, the second produces space-time--time, and so on.  Space-time is a
generalized de Sitter space.  Space-time--time, in which quantum effects show
up, is a happy hybrid of two notable attempts at a unified theory of gravity
and electromagnetism: the Kaluza five-dimensional geometry \cite{3} and the
Weyl conformal geometry \cite{4}.  Brought together in this way those theories
lose their undesirable properties while retaining their useful ones.  That the
space-time--time  geometry both induces quantum effects and includes gravity
(along with other fields) calls into question the rationale behind the search
for a quantum theory of gravity.  This presentation and its author might
therefore be regarded as intruders from a school devoted to ``Escapes from
Quantum Gravity'', conducted in a parallel universe.

The geometric construction in question can be understood initially by reference
to Fig.~1, which shows in cross section two neighboring spheres $S$ and $S'$
in euclidean 3-space $\mathbb E^3$ with centers $C$ and $C'$ the distance $ds$
apart, and with radii $R$ and $R + dR$.  The angle $d\alpha$ in which they
intersect is found, by application of Pythagoras' theorem to the infinitesimal
right triangle in the middle, to be given by
$d\alpha^2 = (1/R^2)(ds^2 - dR^2)$, which shows $d\alpha$ to be the line
element of a metric of diagonal signature $+++\;-$ on the four-dimensional
manifold ${\cal M}^4$ whose points are the 2-spheres of $\mathbb E^3$.  Note
that $d\alpha$ is invariant under conformal transformations of $\mathbb E^3$.
\begin{figure}[t]
\begin{minipage}{3in}
   \includegraphics[bb = 149 430 590 721, width=4.25in]{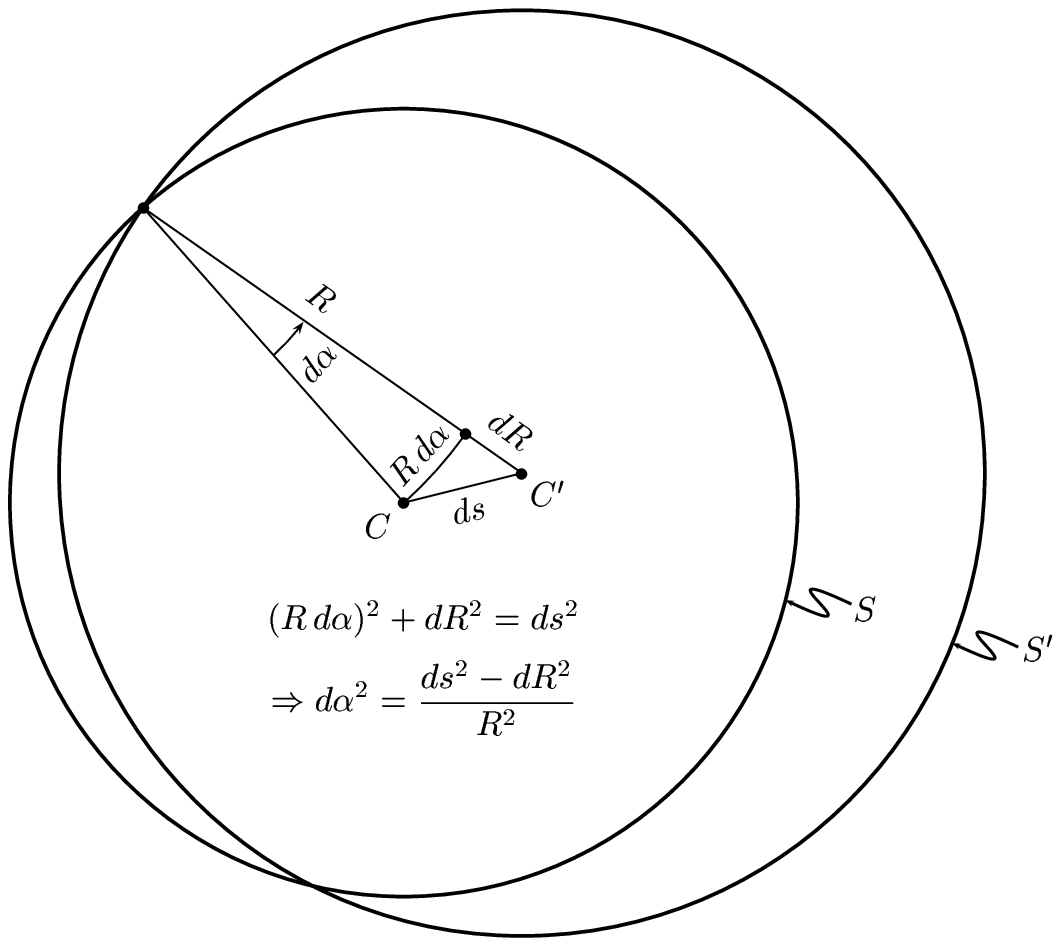}
\vskip -0.15in
   \caption{The `angle' line element.}
\end{minipage}
\hspace{-.35in}
\begin{minipage}{4in}
   \includegraphics[width=4.25in]{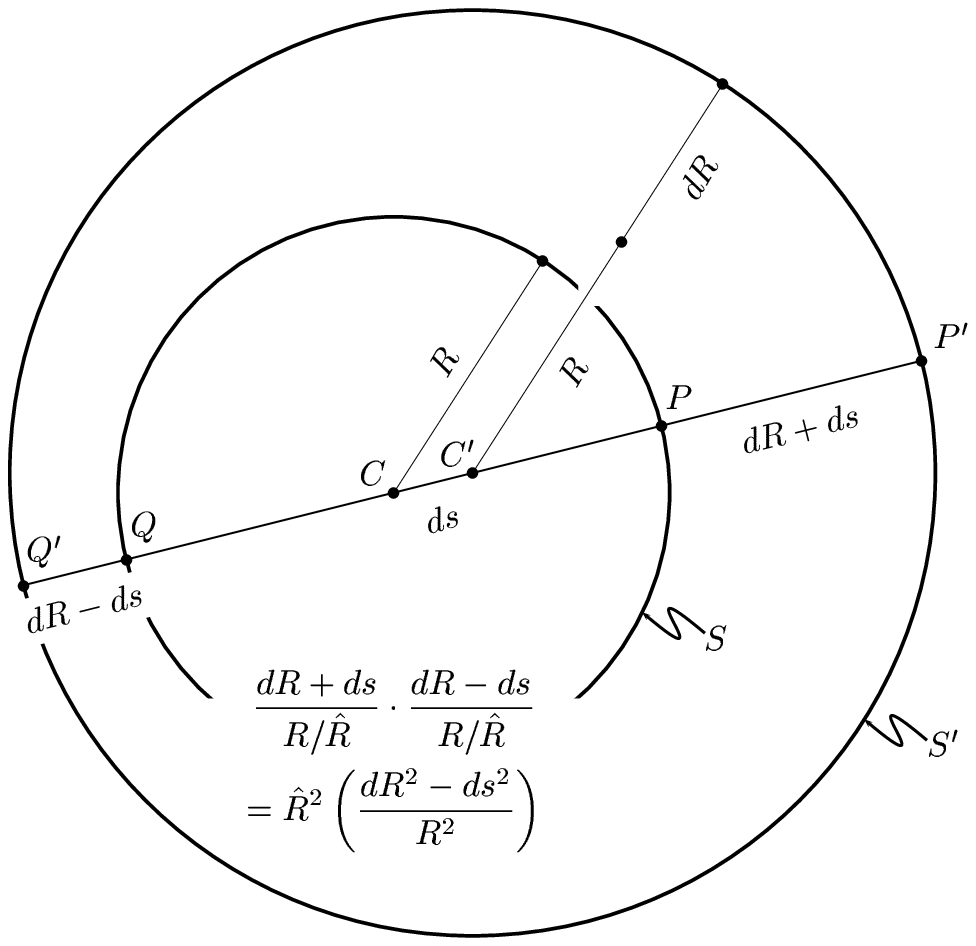}
\vskip -0.15in
   \caption{The `displacement' line element.}
\end{minipage}
\vskip -0.15in
\end{figure}

A more general version of the construction is exhibited in Fig.~2.  Here,
instead of $\mathbb E^3$ there is a three-dimensional manifold ${\cal M}^3$
with a positive-definite riemannian metric $G$, whose geodesic \linebreak
2-spheres are the points of ${\cal M}^4$; $S$ and $S'$ are two such spheres,
$Q'QCC'PP'$ is the geodesic through their centers $C$ and $C'$, and the line
element is generated as the product of the distances (relative to a scaled
radius $R/\Rhat$) by which $P$ and $Q$ are displaced when $S$ is magnified
radially by the factor $1 + dR/R$ and its center $C$ is shifted a distance
$ds$ along the geodesic to $C'$, which in effect converts $S$ to $S'$.  The
line element that results is given by $d\tau^2 = (\Rhat/R)^2(dR^2 - ds^2)$.

Defining a new coordinate $t$ by $t := -\ln (R/\Rhat)$ makes
\begin{equation}
d\tau^2 = \Rhat^2 dt^2 - e^{2t} ds^2,
\end{equation}
a generalization of the de Sitter space-time metric for an empty expanding
universe of uniform radius of curvature $\Rhat$, to which metric it reduces
when ${\cal M}^3 = \mathbb E^3$.  Thus, by means of a simple geometric
construction we have produced from (more aptly, discovered within) the geometry
of three-dimensional space the geometry of a four-dimensional space-time.

The tensor product version of the metric specified by equation (1) is
\begin{equation}
\Ghat = \Rhat^2 (dt \otimes dt) - e^{2t} G
      = \Rhat^2 (dt \otimes dt) - e^{2t} (dx^m \otimes g_{mn} dx^n),
\end{equation}
where $G$ is the metric of ${\cal M}^3$.  For this space-time metric
$\partial_t$ is a `conformal semi-Killing' vector field, in the sense that
${\cal L}_{\partial_t} \Ghat = -2\, e^{2t} G$, where ${\cal L}_{\partial_t}$
denotes Lie differentiation along $\partial_t$.  Observing that
$-e^{2t} G = \Ghat - (\Ghat \partial_t \partial_t)^{-1}
(\Ghat \partial_t \otimes \Ghat \partial_t)$,
one sees that to capture least restrictively in a generic space-time metric
$\Ghat$ on a manifold ${\cal M}^4$ the essence of the geometrical construction
in question it is sufficient to subject $\Ghat$ to the constraint that there
exist on ${\cal M}^4$ a time-like vector field $\xi$ such that
${\cal L}_{\xi}\Ghat =
2\, [\Ghat - (\Ghat \xi \xi)^{-1} (\Ghat \xi \otimes \Ghat \xi)]$.  It is then
easy to see that, in a coordinate system $\dl \, x^m, t \, \dr$ adapted to
$\xi$ so that $\xi = \partial_t$, $\Ghat$ takes the form
\begin{equation}
\eqalign{
\Ghat &= \phi^2 (A + dt) \otimes (A + dt) - e^{2t} G \\
      &= \phi^2 (A_m dx^m + dt) \otimes (A_n dx^n + dt)
            - e^{2t} (dx^m \otimes g_{mn} dx^n),}
\end{equation}
with $\phi$, $A_m$, and $g_{mn}$ independent of $t$.

Having arrived at the space-time geometry described by the metric $\Ghat$ on
the manifold ${\cal M}^4$, we can repeat the construction, applying it this
time to the geodesic 3-spheres of ${\cal M}^4$ to produce a metric on the
five-dimensional manifold ${\cal M}^5$ whose points are those geodesic
\linebreak
3-spheres.  Generalizing that metric in the manner that produced (3) we obtain
\begin{equation}
\eqalign{
\Ghat &= e^{2 \zeta} G 
           + \epshat \phi^2 (A + d \zeta) \otimes (A + d \zeta) \\
      &= e^{2 \zeta} (dx^\mu \otimes g_{\mu \nu} \, dx^\nu)
           + \epshat \phi^2 (A_\mu \, dx^\mu + d\zeta)
                    \otimes (A_\nu \, dx^\nu + d\zeta),}
\end{equation}
referred to a coordinate system $\dl \, x^\mu, \zeta \, \dr$ such that
$\xi = \partial_\zeta$, where now $G$ is the space-time metric and $\phi$,
$A_\mu$, and $g_{\mu \nu}$, the counterparts of the previous $\phi$, $A_m$, and
$g_{mn}$, are independent of $\zeta$.  The factor $\epshat$ enters because
there are two kinds of spheres (hyper-hyperboloids, actually) in space-time.
Those whose points lie in spacelike directions from their centers require that
$\epshat = 1$, those whose points lie in timelike directions from their centers
require $\epshat = -1$.  In either case, because the construction that produced
$\Ghat$ from space-time is the same one that produced space-time from space, it
is justified, indeed unavoidable, to label the manifold ${\cal M}^5$ with the
metric $\Ghat$ a space-time--time, and to identify $\zeta$ as a temporal
coordinate, albeit of a secondary nature distinct from that of the primary time
coordinate $t$.  Ultimately the $\epshat = \pm 1$ distinction should be
resolved by allowing $\zeta$ to be a complex coordinate.  For present purposes,
however, $\zeta$ will be kept real and $\epshat$ will be 1, the points of
${\cal M}^5$ being therefore the spacelike spheres of ${\cal M}^4$.

The geometry of space-time--time is the previously mentioned happy hybrid of
the Kaluza and the Weyl geometries.  As in Kaluza's theory, $A$ is the
space-time covector potential of the electromagnetic field 2-form $F$
defined by $F := -2\, d_\wedge A$.  The coordinate transformation
$\zeta' = \zeta - \lambda$, with $\lambda$ independent of $\zeta$, generates
in one stroke both the electromagnetic gauge transformation $A' = A + d\lambda$
and the Weyl conformal transformation $G' = e^{2 \lambda} G$.  The geodesic
paths of $\Ghat$ are taken to be the histories of test particles.  To analyze
these histories for physical content one introduces the frame system
$\{e_\mu, e_5\} := \{\partial_\mu - A_\mu \, \partial_\zeta,
\phi^{-1} \partial_\zeta\}$ and its dual coframe system
$\{\omega^\mu, \omega^5\} := \{dx^\mu, \phi (A_\mu \, dx^\mu + d\zeta)\}$, for
which $\Ghat = e^{2 \zeta} (\omega^\mu \otimes g_{\mu \nu} \, \omega^\nu) +
\omega^5 \otimes \omega^5$ and $e_5$ is orthogonal to the $e_\mu$.  The
velocity $\pdot$ of a path $p\!: \mathbb R \to {\cal M}^5$ then is expressed by
$\pdot = \pdot^\mu e_\mu + \pdot^5 e_5$, and one can introduce the following
definitions for a test particle whose history is $p$:
\begin{eqnarray}
    P\!\!\!\!&:= \Ghat \pdot
               = (e^{2 \zeta} \pdot^\kappa g_{\kappa \mu}) \omega^\mu
                    + \pdot^5 \omega^5
               =: P_\mu \omega^\mu + P_5 \omega^5 \qquad
                  &\mbox{(momentum),} \\
  \mo\!\!\!\!&:= (P_\mu \, g^{\mu \nu} P_\nu)^{1/2}
               = e^{2 \zeta} (\pdot^\mu g_{\mu \nu} \pdot^\nu)^{1/2}
                   \hspace{79pt}
                  &\mbox{(rest mass),} \\
    q\!\!\!\!&:= \phi P^5 = \phi \pdot^5 = \phi^2 (A_\mu \pdot^\mu + \zetadot)
                   \hspace{101pt}
                  &\mbox{(electric charge),} \\
 \tau\!\!\!\!&:= \int (\pdot^\mu g_{\mu \nu} \pdot^\nu)^{1/2}
               = \int e^{-2 \zeta} \mo
                   \hspace{110pt}
                  &\mbox{(proper time),} \\
u^\mu\!\!\!\!&:= \displaystyle{\frac{dp^\mu}{d\tau}
              := \frac{\pdot^\mu}{\taudot}}
                   \hspace{179pt}
                  &\mbox{(proper velocity).}
\end{eqnarray}
Of these $P$ and $q$ are gauge-invariant and the others are not, although the
condition $\mo = 0$ is gauge-invariant.

Taking $p$ to be a secondarily timelike geodesic parametrized by arclength,
that also is primarily timelike in that $\pdot^\mu g_{\mu \nu} \pdot^\nu > 0$,
one has that $|\pdot|^2 = \Ghat(p) \pdot \pdot =
e^{2 \zeta} (\pdot^\mu g_{\mu \nu} \pdot^\nu) + (\pdot^5)^2 =
e^{-2 \zeta} \mo^2 + (q/\phi)^2 = 1$ and that
$\pbdotdot = \pbdotdot^\mu e_\mu + \pbdotdot^5 e_5 = 0$.  The equation
$\pbdotdot^\mu = 0$ is equivalent to
\begin{equation}
\frac{d(\mo u^\mu)}{d \tau} +
(\mo u^\kappa) \Gamma_\kappa {}^\mu {}_\lambda u^\lambda
       = q F^\mu {}_\lambda u^\lambda - \mo A^\mu
           + e^{2 \zeta} \frac{(q/\phi)^2}{\mo} (\ln \phi)^{.\mu},
\end{equation}
and $\pbdotdot^5 = 0$ is equivalent to
\begin{eqnarray}
\qdot = e^{-2 \zeta} \mo^2 = 1 - (q/\phi)^2,
 & \qquad \mbox{also to} \qquad &
\frac{dq}{d\tau} = \mo = e^\zeta [1 - (q/\phi)^2]^{1/2}.
\end{eqnarray}
Equations (11) and (7), together with $e^{-2 \zeta} \mo^2 + (q/\phi)^2 = 1$,
entail that
\begin{equation}
(\mo^2)\dot{}\,
  = 2 [-\mo^2 A_\mu + e^{2 \zeta} (q/\phi)^2 (\ln \phi)_{.\mu}] {\pdot}^\mu.
\end{equation}

Equation~(10) shows the rate of change of the particle's space-time momentum
$\mo u^\mu$ with respect to its proper time to be governed by four `forces':
the Einstein--Newton force
$-(\mo u^\kappa) \Gamma_\kappa {}^\mu {}_\lambda u^\lambda$, the Lorentz force
$q F^\mu {}_\lambda u^\lambda$, the `Weyl force' $-\mo A^\mu$, and the `Kaluza
force' $(\mo)^{-1} (q/\phi)^2 (\ln \phi)^{.\mu}$.  Equation~(11) produces
generic behavior of $q$ that is exemplified in the solution $q(\tauhat) =
\phi \tanh (\tauhat/\phi)$, where $\tauhat$ is secondary proper time and $\phi$
is taken to be constant.  From $e^{-2 \zeta} \mo^2 + (q/\phi)^2 = 1$ follows
$\zeta(\tauhat) = \ln (\mo(\tauhat) \cosh (\tauhat/\phi))$.  If also
$A_\mu = 0$, then equation (12) tells that $\mo$ is constant, and (8) yields
$\tau(\tauhat) = \mo \int e^{-2 \zeta (\tauhat)}\, d\tauhat =
(\phi/\mo) \tanh (\tauhat/\phi)$, which shows the particle's proper lifetime to
be confined to the open interval $(\tau(-\infty), \tau(\infty))$
($= (-\phi/\mo, \phi/\mo)$), thus associates longer lifetimes with smaller rest
masses, shorter lifetimes with larger rest masses.  Noting in passing the
equivalence of inertial mass and passive gravitational mass implied by the two
appearances of $\mo$ in the lefthand member of (10), one divides by $\mo$ to
obtain $du^\mu/d\tau + u^\kappa \Gamma_\kappa {}^\mu {}_\lambda u^\lambda = 0$,
which says that the particle's space-time track is a geodesic of the Einstein
geometry.  The particle's coordinates $x^\mu$ will in general have $\tauhat$
dependence similar to that of $\tau$, in consequence of which the particle's
space-time track will have an endpoint event ${\cal E}_1$ at which it appears
suddenly, traveling with velocity $(dx^m/dt)(-\infty)$ (if $dt/d\tau > 0$), and
an endpoint event ${\cal E}_2$ at which it disappears just as suddenly,
traveling with velocity $(dx^m/dt)(\infty)$.

Essential features of this behavior will persist in the generic case where
$\phi$ and $A_\mu$ are not restricted.  In particular, the space-time track
will end at events ${\cal E}_1$ and ${\cal E}_2$, and $q$ will grow from
$-\phi({\cal E}_1)$ to 0 and on to $\phi({\cal E}_2)$ and $\zeta$ will depart
from and return to $\infty$, as $\tauhat$ goes from $-\infty$ to $\infty$.
Figure~3 is a schematic representation of a geodesic exhibiting such behavior.
\vskip 10pt
\begin{figure}[ht]
   \includegraphics[bb = -278 548 195 893, width=5.5in]{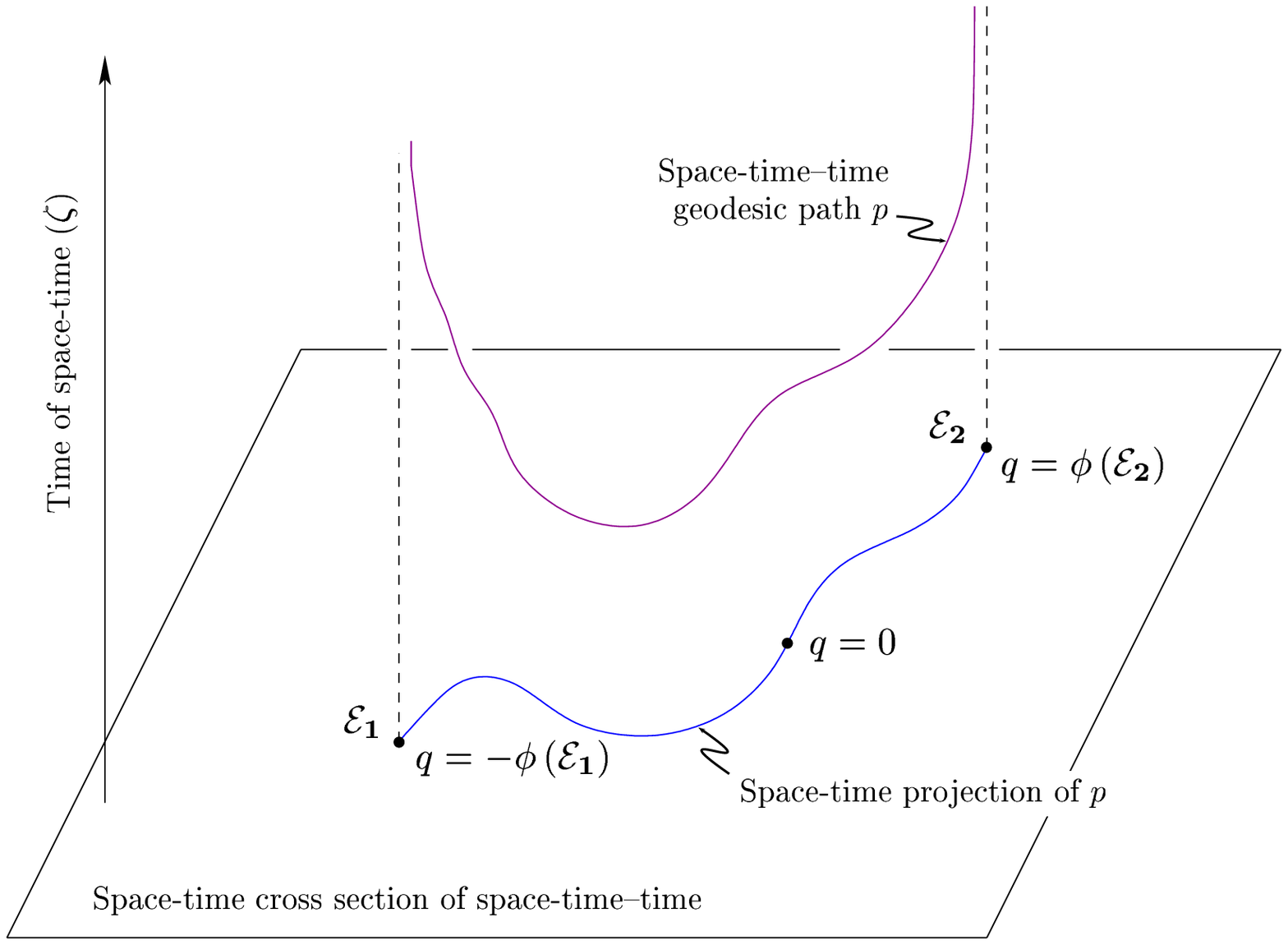}
   \caption{A space-time--time geodesic~and~its space-time projection.}
\end{figure}

The variational principle
$0 = \delta \int_{{\cal D} \times [a,b]} (\Psihat - \Psihat_\infty) \, d\Vhat$,
where $\Psihat$ is the curvature scalar of $\Ghat$, given by
\begin{equation}
\eqalign{
\Psihat &= e^{-2 \zeta} \Psi + 6 e^{-2 \zeta} \left[ A^\kappa{}_{: \kappa}
                             - A^\kappa A_\kappa \right] \\
        & \qquad - 2 e^{-2 \zeta} \phi^{-1}
                                  \left[ \phi^{. \kappa}{}_{: \kappa}
                 - 2 A^\kappa\phi_{. \kappa} \right]
                 + (1/4) e^{-4 \zeta} \phi^2
                                      F_\kappa{}^\lambda F_\lambda{}^\kappa
                 - 20 \, \phi^{-2},}
\end{equation}
$\Psihat_\infty := \lim_{\zeta \to \infty} \Psihat = -20 \, \phi^{-2}$, and
$\cal D$ is a region of space-time, produces the following field equations,
obtained by varying $\phi$ and $A_\mu$, respectively, where
$k := 2(b - a)/(e^{2b} - e^{2a})$:
\begin{equation}
A^\kappa{}_{: \kappa} - 3 A^\kappa A_\kappa
 = - (3/8) k \phi^2 F_\kappa{}^\lambda F_\lambda{}^\kappa - (1/2) \Psi
\end{equation}
and
\begin{equation}
F^{\mu \kappa}{}_{: \kappa} + 3 F^{\mu \kappa} (\ln \phi)_{. \kappa}
 = -2 k^{-1} \phi^{-2} \left[ (\ln \phi)^{. \mu} + 6 A^\mu \right].
\end{equation}

In \cite{5} the analogous equations for space-time are shown to have
spherically symmetric solutions of the `traversable wormhole' type, similar to
those in \cite{6}.  In the space-time--time case, with $G$ describing a
nongravitating, static, spherically symmetric, traversable wormhole, with
$A = V(r) \, dt$, and with $\ln \phi = U(r)$, numerical integration yields
a variety of solutions for which, as $r \to \infty$, $V(r)$ is asymptotic to
a Coulomb potential $Q/r$.  For one of these, typical of a large class, $U(r)$
has, for $r \geq 0$, the shape shown in Fig.~4.  The spacing of successive
bottoms of the potential wells of $U(r)$, located at $r = r_n$,
$n = 1,2,3,\cdots$, is asymptotic to $2n$, consistent with $r_n$'s growing
asymptotically as $n^2$.  Figure~5 is a graph of an artificial version
$\bar U(r)$ of $\ln \phi$ with $r_n = n^2$ and potential wells of uniform
depth, to be used for illustrative purposes.
\begin{figure}[h]
\begin{minipage}{2.9in}
   \includegraphics[bb = 86 -384 566 -90, width=2.9in]{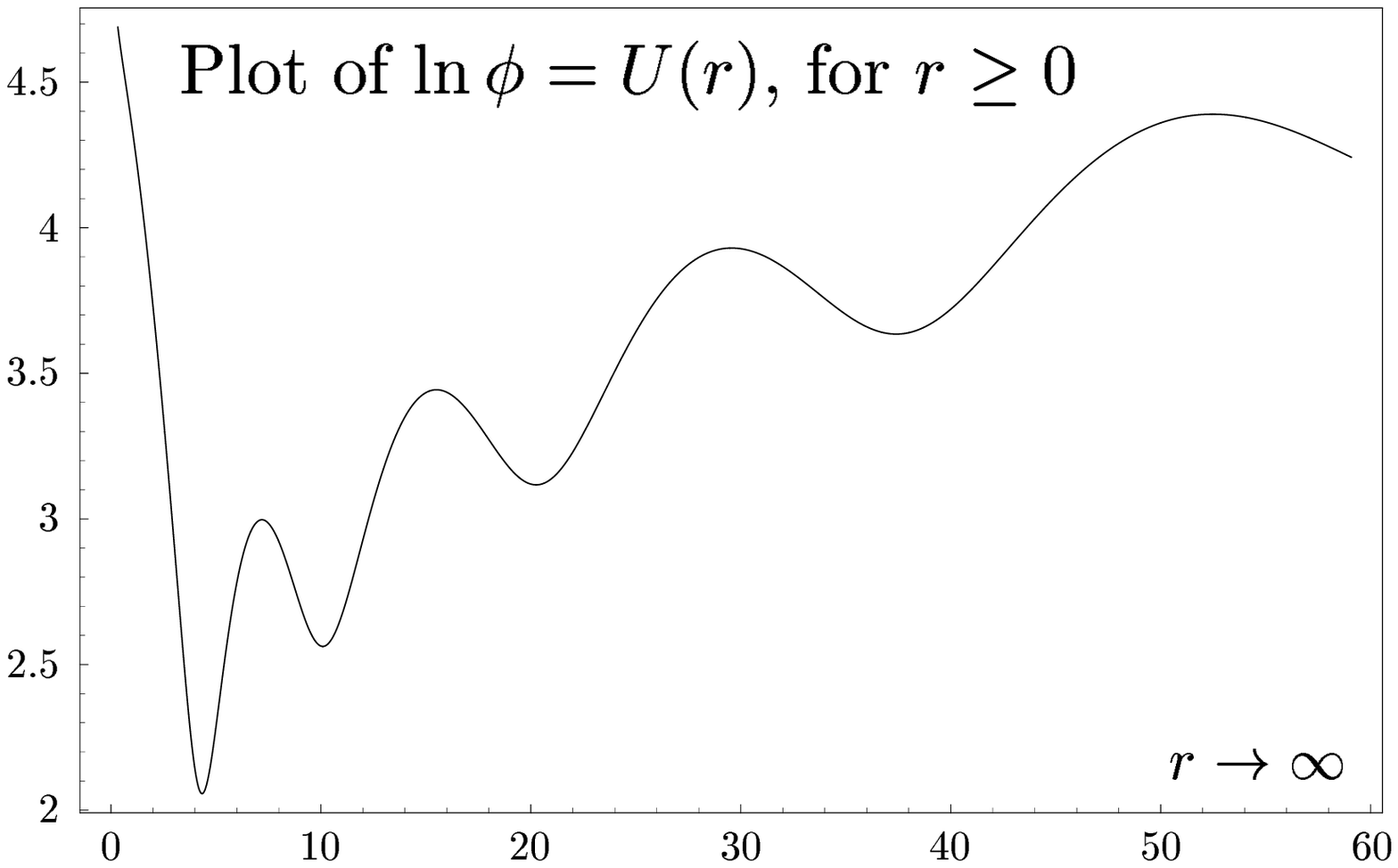}
   \caption{A computed potential.}
\end{minipage}
\hspace{0.2in}
\begin{minipage}{2.9in}
   \includegraphics[bb = 86 -384 566 -90, width=2.9in]{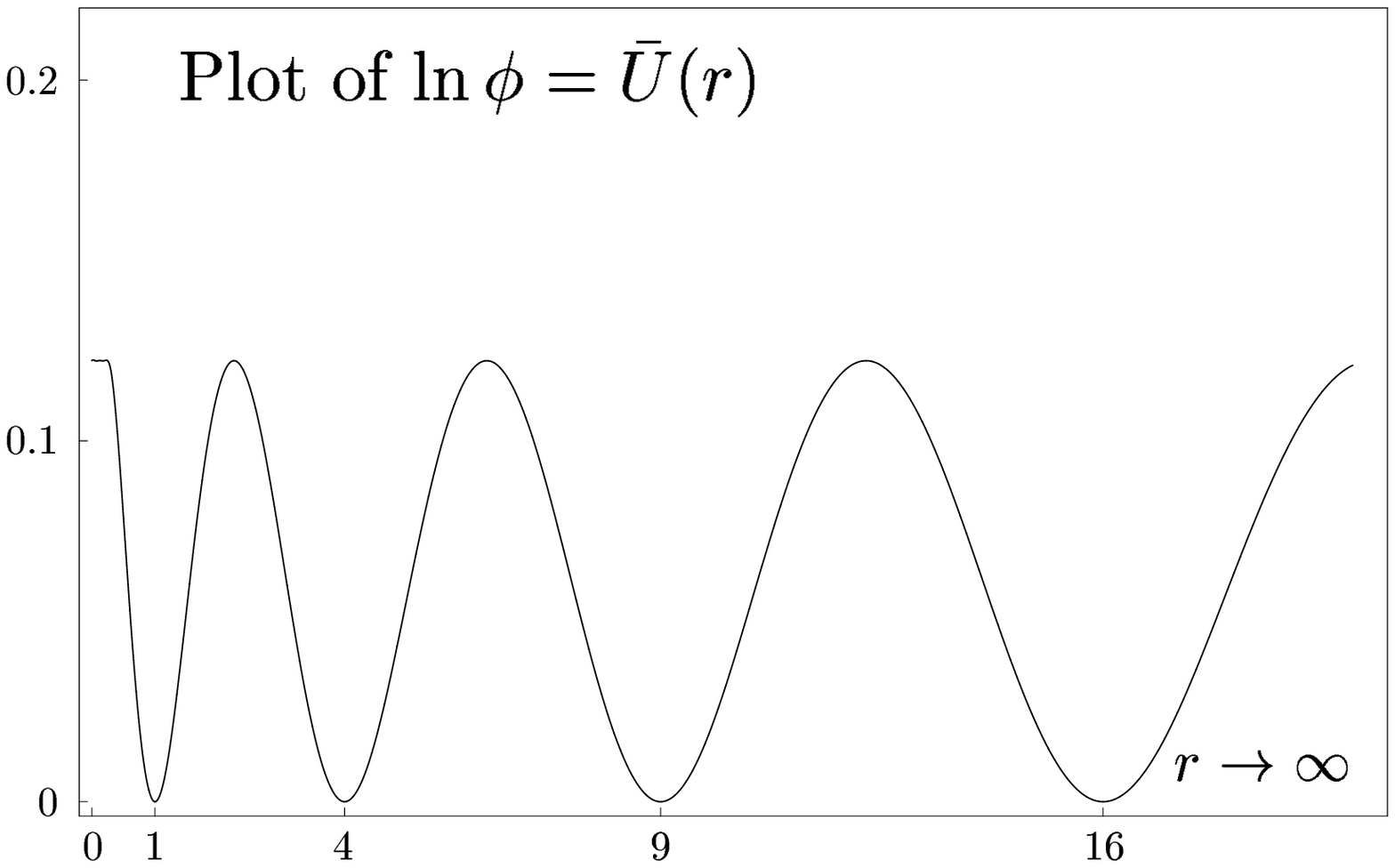}
   \caption{An artificial potential.}
\end{minipage}
\end{figure}

Now comes a most remarkable aspect of a test particle's space-time behavior:
both as $\tauhat \to -\infty$ and as $\tauhat \to \infty$ the factor
$e^{2 \zeta}$ that in equation (10) couples the Kaluza force to the momentum
rate becomes infinite, which causes that force to {\em infinitely} dominate the
other three, and to push each of the terminal events ${\cal E}_1$ and
${\cal E}_2$ toward a bottom of one of the potential wells of $\ln \phi$.  Thus
a scatter plot of ${\cal E}_1$ and ${\cal E}_2$ generated by random choices of
initial conditions $p(0)$ and $\pdot(0)$ for the test particle's path $p$ would
show high densities near those potential well bottoms, low densities elsewhere.
Figure~6 illustrates this behavior, which clearly suggests the possibility of a 
(for the present, only qualitative) deterministic underpinning of quantum
mechanics like that provided to thermodynamics by the molecular theory of
gases.

As illustrated in Fig.~3, neither of the events ${\cal E}_1$ and ${\cal E}_2$
is a projection of a point on the geodesic path $p$: they are only limits of
such projections as $\tauhat \to \pm \infty$ and $\zeta \to \infty$.  This
suggests that the particle whose space-time track the projection is exists
neither at or before ${\cal E}_1$ nor at or after ${\cal E}_2$, rather exists
only between ${\cal E}_1$ and ${\cal E}_2$.  Think back, however, to the
original conception of events in space-time as 2-spheres in space, with
$t = -\ln (R/\Rhat)$.  Under this interpretation the projection of $p$ is a
one-parameter family of spheres which converges at either of these
events~$\cal E$ to the sphere $S({\cal E})$ centered at the spatial location of
\begin{figure}[t]
  \includegraphics[bb= -225 290 225 580, width=6.25in]{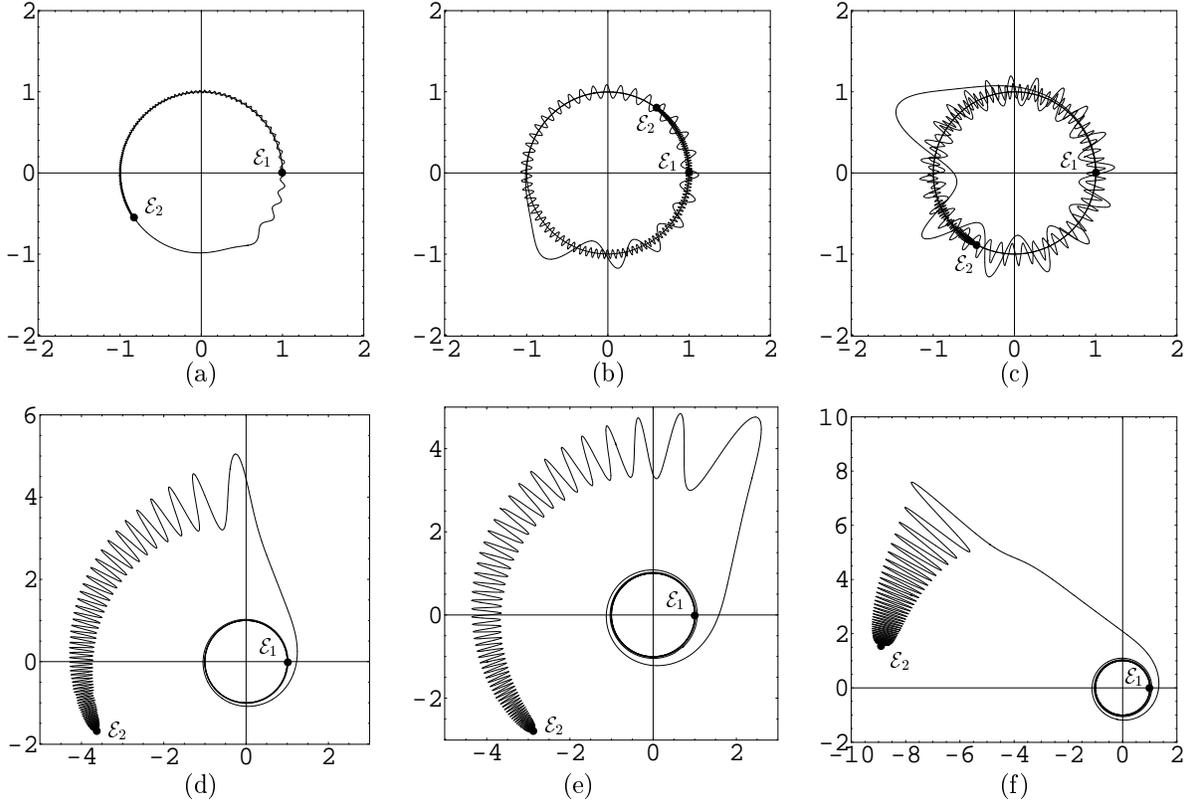}
\vspace{10pt}
  \caption{Sample tracks of test particles following
space-time--time geodesics governed by the artificial potential
$\ln \phi = \bar U(r)$ of Fig.~5 and a Coulomb potential $\bar V(r) = Q/r$.
The particles appear at ${\cal E}_1$, follow the track in the counterclockwise
direction, and disappear at ${\cal E}_2$.
In each case ${\cal E}_1$ is at $r = 1$, $\varphi = 0$, with
$q = -\phi({\cal E}_1) = -1$ and $\mo = 1$.  The initial angular velocities
$d\varphi/dt$ are: (a)~0.0050; (b)~0.0101; (c)~0.0151; (d)~0.0352; (e)~0.0502;
(f)~0.0602. The numbers of complete revolutions in the orbits, and the
locations of ${\cal E}_2$, are
(a)~1, $r = 0.995545$, $\varphi = 212\,\degree$\,;
(b)~3, $r = 0.992591$, $\varphi = \;51\,\degree$\,;
(c)~4, $r = 0.993775$, $\varphi = 240\,\degree$\,;
(d)~5, $r = 4.00752$, $\;\varphi = 204\,\degree$\,; \linebreak
(e)~7, $r = 3.97478$, $\;\varphi = 224\,\degree$\,;
(f)~8, $r = 9.09021$, $\;\varphi = 170\,\degree$\,.
In each case at the end~$q = \phi({\cal E}_2) \approx 1$.  Such deterministic
geodesics of space-time--time can model (qualitatively, at least) quantum
behavior of electrons in atoms, of `quantum jumps' between electron orbits in
particular.}
\vspace{-5pt}
\end{figure}
$\cal E$ and with the nonzero radius~$\Rhat e^{-t({\cal E})}$.  By definition
$\cal E = S(\cal E)$, and therefore every test particle track that has $\cal E$
as one of its endpoints would include $S({\cal E})$ by continuity.
What is more, the full geodesic path is itself a one-parameter family of
space-time--time points, thus of (hyper-hyperboloidal) `spheres' of space-time.
The 3-`sphere' $S_\zeta$ that is the point at $\dl \, x^\mu, \zeta \, \dr$ has
as its center the spatial 2-sphere that is the event at $\dl \, x^\mu \, \dr$
in space-time.  The radius of $S_\zeta$ is $\Rhat e^{-\zeta}$, which goes to
zero at $\cal E$.  According to the space-time metric constructed in
Fig.~1, $S_\zeta$ is the set of all 2-spheres that lie an angular distance
$\Rhat e^{-\zeta}$ from the central 2-sphere.  This set is a union of disjoint
subsets each of which is a one-parameter family of 2-spheres all mutually
tangent to one another at a single point of the central 2-sphere, which they
all intersect in an angle of radian measure $\Rhat e^{-\zeta}$.  These
families, which in the conventional sense are null generators of the
hyper-hyperboloid that is the 3-sphere $S_\zeta$, are null geodesics of
space-time which broadcast the location and size of the central 2-sphere, both
forward in time and backward.  As $\zeta \to \infty$ the 2-spheres all become
tangent to the central sphere.  Conventionally put, the hyperboloid $S_\zeta$
collapses to a null cone, whose vertex is the event corresponding to (i.\,e.,
equal to by definition) the limiting central sphere $S({\cal E})$.  The events
on or interior to the past null cone of ${\cal E}_1$ and the future null cone
of ${\cal E}_2$ receive no information about the particle, but every other
event is notified of the particle's existence (by being an event on a null
generator of one of the hyperboloids $S_\zeta$ in the space-time--time
geodesic, thus by being a 2-sphere in one of the families of mutually tangent
2-spheres whose union is $S_\zeta$).  If we now consider the test particle
following the path $p$ to in fact {\em be} a 2-sphere in space, going forward
in (primary) time by shrinking, then, because it has nonzero spatial radius at
${\cal E}_1$ and ${\cal E}_2$, the particle can be deemed to exist there, even
though it is visible only between ${\cal E}_1$ and ${\cal E}_2$.

A second test particle whose track shared with that of the first an endpoint
$\cal E$ would have in common with the first the 2-sphere $S({\cal E})$, thus
would be, for an instant at least, the same particle.  The two particles could
be thought of as extensions of one another, as well as of all other particles
that shared the endpoint $\cal E$.  Such an event would be an `interaction'
event, not unlike a vertex in a quantum mechanical Feynman diagram.  Built in
to the interaction would be that all participating charged particles have the
same charge magnitude $|q| = \phi({\cal E})$.

There is much left unreported here, but I trust that what has been reported is
sufficient to lend credence to the proposition that some measure of the physics
of quantum phenomena can be extracted, qualitatively and to some degree
quantitatively, from the geometry of three-dimensional space.

\end{document}